\documentclass[doublecol]{epl2} % for 2 columns style with line numbers
\title{Effects of Vanadium doping on BaFe$_{2}$As$_{2}$}
\shorttitle{}

\author{Xing-Guang Li\inst{1}\footnote{These authors contributed equally to this paper.} \and Jie-Ming Sheng\inst{1,2}\footnotemark[1] \and Cong-Kuan Tian\inst{1} \and Yi-Yan Wang\inst{1} \and Tian-Long Xia\inst{1} \and Le Wang\inst{1} \and Feng Ye\inst{2} \and Wei Tian\inst{2} \and Jin-Chen Wang\inst{1} \and Juan-Juan Liu\inst{1} \and Hong-Xia Zhang\inst{1} \and Wei Bao\inst{1} \and Peng Cheng\inst{1}\footnote{Corresponding author. E-mail: pcheng@ruc.edu.cn}}
\shortauthor{F. Author \etal}

\institute{
  \inst{1}  Department of Physics and Beijing Key Laboratory of Opto-electronic Functional
Materials \& Micro-nano Devices, Renmin University, Beijing
100872, China
\\
  \inst{2} Neutron Scattering Division, Oak Ridge National Laboratory, Oak Ridge, TN 37831, USA
}

\pacs{74.70.Xa}{Pnictides and chalcogenides}
\pacs{74.62.Dh}{Effects of crystal defects, doping and
substitution}

\abstract{We report an investigation of the structural, magnetic
and electronic properties of Ba(Fe$_{1-x}$V$_{x}$)$_{2}$As$_{2}$
using x-ray, transport, magnetic susceptibility and neutron
scattering measurements. The vanadium substitutions in Fe sites
are possible up to $\sim$40\%. Hall effect measurements indicate
strong hole-doping effect through V doping, while no
superconductivity is observed in all samples down to 2~K. The
antiferromagnetic and structural transition temperature of
BaFe$_{2}$As$_{2}$ is gradually suppressed to finite temperature
then vanishes at x=0.245 with the emergence of spin glass
behavior, suggesting an avoided quantum critical point (QCP). Our
results demonstrate that the avoided QCP and spin glass state
which were previously reported in the superconducting phase of
Co/Ni-doped BaFe$_{2}$As$_{2}$ can also be realized in
non-superconducting Ba(Fe$_{1-x}$V$_{x}$)$_{2}$As$_{2}$.}

\begin{document}
\maketitle
\section{Introduction}

In both cuprate and Fe-based high-T$_c$ superconductors,
investigation of the interplay between the magnetism and
superconductivity through chemical doping is critical to explore
the superconducting mechanism\cite{Cuprate,Hosono}. In the
iron-pnictides phase diagram, besides superconductivity and
antiferromagnetic (AFM) phase, various exotic states including
incommensurate short-range AFM order, G-type AFM order, C$_4$
magnetic phase and spin glass state were observed which reveals
the rich physics controlled by the interplay between the spin,
charge and orbital degrees of freedom in this
system\cite{FeReview,IC_Ni,Cr,Cr_Neutron,Mn,Ni_QCP,C4,Co_Glass,Ni_Glass}.

For the well-known FeAs-122 parent compound BaFe$_{2}$As$_{2}$,
either electron doping (e.g., Co, Ni, Cu, Rh, Ir, Pt,
Pd)\cite{Co_Sc,NiCu_Sc,RhIrPd_Sc,Pt_Sc} or isoelectronic doping
(e.g., Ru, P)\cite{Ru_Sc,P_Sc} in the FeAs plane could easily
instigate superconductivity. On the hole doping side, although
alkali metal doping in the Ba site could induce superconductivity
with transition temperature as high as 38 K\cite{K122}, attempts
of hole doping in the FeAs plane (e.g., Mn, Cr, Mo)\cite{Mn,Cr,Mo}
show that the stripe AFM and structural transitions are suppressed
to some extent but superconductivity was never observed. This
contrasting behavior has not been well understood. Furthermore,
the discovery of a novel C$_4$ magnetic phase only in the
hole-doped 122 materials has drawn a lot of research
interests\cite{C4}. Therefore more investigations are needed for
the hole doping cases in the FeAs plane.

In this paper we report the effect of V-doping on
BaFe$_{2}$As$_{2}$ in a bulk property and neutron/x-ray
diffraction study. The hall effect measurements indicate that V
substitution for Fe results in hole doping. Although
superconductivity is not observed in
Ba(Fe$_{1-x}$V$_{x}$)$_{2}$As$_{2}$ as in other in-plane
hole-doping cases, an avoided Quantum critical point with the
emergence of spin glass state are found in heavily doped samples.

\section{Experiment}
Single crystals of Ba(Fe$_{1-x}$V$_{x}$)$_{2}$As$_{2}$ were grown
by FeAs/VAs self-flux method similar to
Ba(Fe$_{1-x}$Co$_{x}$)$_{2}$As$_{2}$\cite{FL}. Polycrystalline
samples were synthesized by heating stoichiometric Barium, FeAs
and VAs in an evacuated quartz tube at 1223~K for 30~h. The
products were reground and annealed at the same temperature twice
to ensure the phase purity. The elemental doping composition was
measured by energy dispersive xray spectroscopy (EDS, Oxford X-Max
50). X-ray diffraction (XRD) patterns of powder and single
crystals were collected from a Bruker D8 Advance x-ray
diffractometer using Cu K$_{\alpha}$ radiation. Resistivity
measurements were performed on a Quantum Design physical property
measurement system (QD PPMS-14T). Magnetization measurement was
carried out in Quantum Design MPMS3. Elastic neutron scattering
measurements were performed using the HB1A triple-axis
spectrometer at HFIR with incident energy E$_{i}$=14.65 meV and
collimation of 40'-40'-40'-80'.

\begin{table}
\caption{Chemical composition analysis of
Ba(Fe$_{1-x}$V$_{x}$)$_{2}$As$_{2}$.} \label{tab.1}
\begin{center}
\begin{tabular}{lcr}
Sample type & Nominal x  & EDS x \\
\hline Single crystal & 0.06 & 0.016 \\
Single crystal & 0.1 & 0.031 \\
Single crystal & 0.14 & 0.051 \\
Single crystal & 0.20 & 0.081 \\
Single crystal & 0.30 & 0.130 \\
Single crystal & 0.40 & 0.183 \\
Polycrystalline & 0.26 & 0.198 \\
Polycrystalline & 0.29 & 0.221 \\
Polycrystalline & 0.30 & 0.245 \\
Polycrystalline & 0.40 & 0.315 \\
Polycrystalline & 0.50 & 0.397 \\
\end{tabular}
\end{center}
\end{table}

\section{Results and discussion}

\begin{figure}
\begin{center}
\includegraphics[width=8cm]{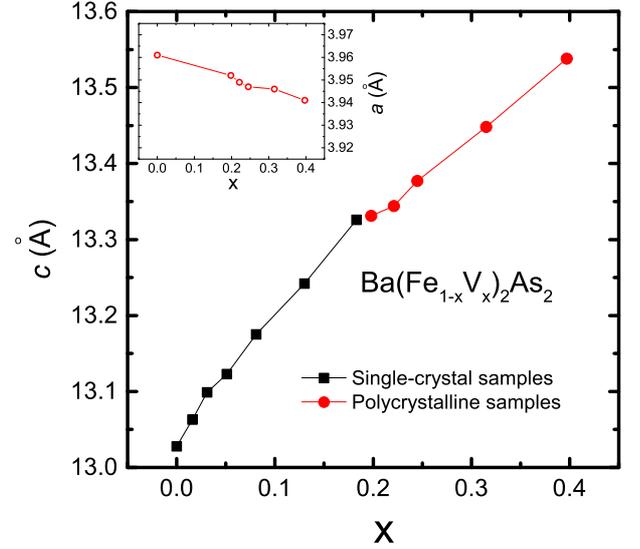}
\caption {Room-temperature lattice parameters as a function of
V-doping value $x$ determined from EDS for
Ba(Fe$_{1-x}$V$_{x}$)$_{2}$As$_{2}$.} \label{Fig.1}
\end{center}
\end{figure}

Single crystals with centimeter or millimeter size are obtained
through self-flux crystal growth method up to nominal doping
x=40\%. It is not possible to grow samples in single crystal form
at a higher doping level. However polycrystalline samples of
Ba(Fe$_{1-x}$V$_{x}$)$_{2}$As$_{2}$ with nominal doping up to
x=50\% can be synthesized with only a small amount of impurity
phases (impurity phases including VAs are estimated to be lower
than 5\% from x-ray diffraction refinement). We characterized all
samples with EDS and the results are listed in Table.1. Figure 1
plots the lattice parameters as a function of V concentration. The
$c$-lattice parameter increases monotonically with increasing $x$
which is the typical feature in hole-doped iron-pnictides
material\cite{122_Review}. Comparing with other hole-doping cases
(Mn, Cr, Mo or K)\cite{122_Review,Kdoping}, Vanadium substitution
effectively expands the $c$-axis, the parameter increases 4\% at
x=0.397. On the other hand, there is a slight contraction for the
$a$-axis lattice parameter which decreases 0.5\% at x=0.397.

\begin{figure}
\begin{center}
\includegraphics[width=8cm]{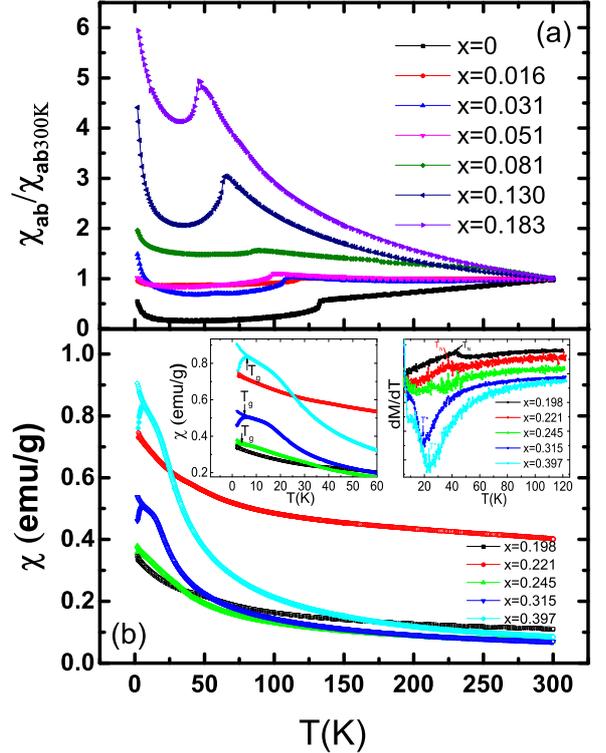}
\caption {The magnetization data for
Ba(Fe$_{1-x}$V$_{x}$)$_{2}$As$_{2}$ in a field $\mu_{0}H$=1~T. (a)
Single crystals with 0$\leq$ x $\leq$0.183 measured along the
ab-crystal direction. (b) Polycrystalline samples with 0.198$\leq$
x $\leq$0.397. The solid symbols represent the ZFC curve and the
open symbols represent the FC curve. The left inset is enlarged
view for T$<$60~K and the right inset shows the temperature
dependence of dM/dT.} \label{Fig.2}
\end{center}
\end{figure}

Magnetization results for Ba(Fe$_{1-x}$V$_{x}$)$_{2}$As$_{2}$
(0$\leq$ x $\leq$0.183) single crystals are presented in Figure
2(a). The data were measured in zero-field-cooling (ZFC) with
magnetic field of 1~T applied parallel to the ab-plane and
normalized by the susceptibility at 300~K. For BaFe$_{2}$As$_{2}$,
the susceptibility decreases linearly with decreasing temperature,
then drops abruptly below the AFM transition with
T$_N$$\approx$133~K. With V-doping, the susceptibility increases
with decreasing temperature and approaches a Curie-Weiss-like
behavior. The AFM transition remains sharp in all doped single
crystals. Figure 2(b) shows both the ZFC and FC magnetization
results for polycrystalline samples at higher doping levels
(0.198$\leq$ x $\leq$0.397). For x=0.198 and x=0.221, the weak AFM
transitions in the magnetic susceptibility is evident from the
dM/dT curves with T$_N$$\approx$43~K and 37~K respectively (inset
of Figure 2(b)). For x=0.245, No sign of magnetic transition can
be inferred from the M-T curve, while the separation between ZFC
and FC susceptibilities at low temperatures indicates a spin glass
behavior (SG). Such behavior becomes even more apparent for
x=0.315 and x=0.397 samples. Furthermore, for x=0.315 and x=0.397
there are new AFM-like kinks (marked as T*) in the M-T curves at
around 20~K, indicating possible new short-range AFM fluctuations.
In addition, we calculated the paramagnetic moments from the
Curie-Weiss fit of susceptibility data for heavily V-doped samples
(x=0.315 and x=0.397). Assuming all the moment contribution is
from V, the results show that the average moment brought by one V
ion is about 2.9$\mu_B$, which is quite close to the theoretical
moment value of one free V$^{3+}$ ion.

\begin{figure}
\begin{center}
\includegraphics[width=8cm]{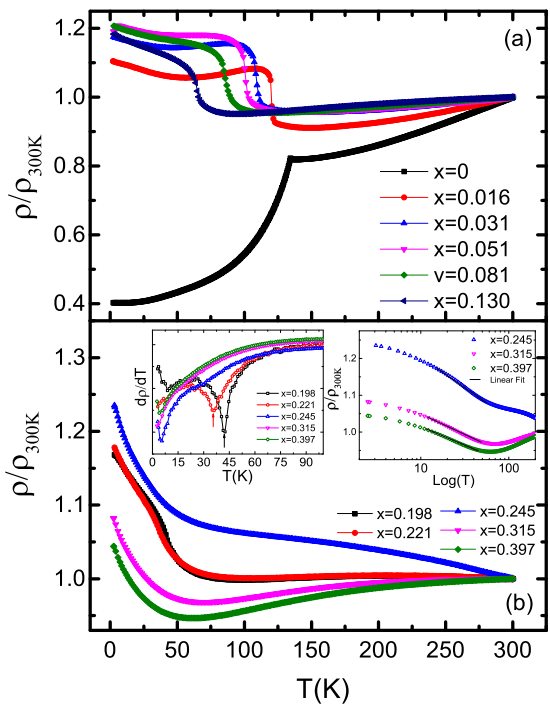}
\caption {Temperature dependence of electrical resistivity for
Ba(Fe$_{1-x}$V$_{x}$)$_{2}$As$_{2}$. (a) Single crystal data with
 0$\leq$ x $\leq$0.13 measured
along the $ab$ direction. (b) Polycrystalline data with
0.198$\leq$ x $\leq$0.397. Left inset shows the T-dependence of
d$\rho$/dT and right inset shows the log(T)-dependence of
resistivity.} \label{Fig.3}
\end{center}
\end{figure}

Figure 3 presents the normalized electrical transport data
$\rho/\rho_{300~K}$. The temperature dependent resistivity for
BaFe$_{2}$As$_{2}$ shows a sharp drop anomaly at the
AFM/structural transition, while for the V-doped samples the
feature becomes a sharp upturn and gradually moves to lower
temperatures with increasing $x$. The temperatures of resistivity
anomalies determined from the minimum of d$\rho$/dT are consistent
with the T$_N$ from the M-T curves. At x=0.198 and x=0.221, the
weak features of resistivity anomalies are still visible from the
d$\rho$/dT curves as shown in the inset of Figure 3(b). However
from x$\geq$0.245, such anomaly could not be detected anymore,
indicating the absence of AFM/structural transition. Furthermore,
the resistivity increases approximately logarithmically with
decreasing temperature below 50~K as shown in the right inset of
Figure 3(b), which is a characteristic of Kondo scattering. This
behavior suggests that the Vanadium dopants act as magnetic
impurities.

\begin{figure}
\begin{center}
\includegraphics[width=8cm]{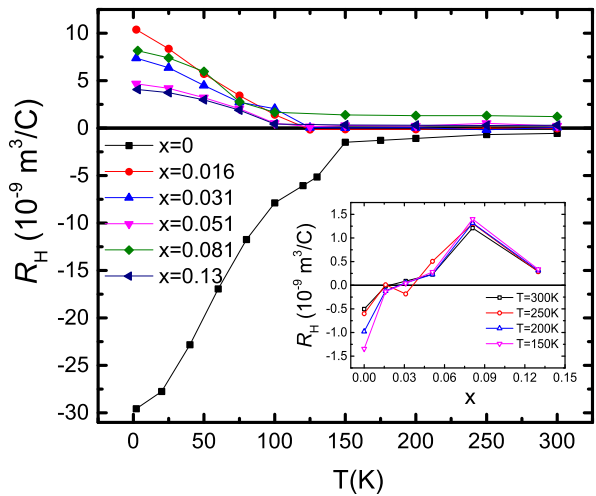}
\caption {The temperature dependence of Hall coefficient R$_H$ for
Ba(Fe$_{1-x}$V$_{x}$)$_{2}$As$_{2}$ (0$\leq$ x $\leq$0.13). The
inset shows doping dependence of R$_H$ at different temperatures.
} \label{Fig.4}
\end{center}
\end{figure}

The Hall coefficients R$_H$ at different temperatures for
Ba(Fe$_{1-x}$V$_{x}$)$_{2}$As$_{2}$ (0$\leq$ x $\leq$0.13) single
crystals are presented in figure 4. R$_H$ changes suddenly from a
negative value in the undoped sample to a positive one with slight
V-doping at low temperatures, and becomes completely positive in
all temperature region measured at x=0.051. Comparing to the Hall
effect in Cr-doping\cite{Cr}, the sign-reversal of R$_H$ occurs at
a much lower doping level indicates that V acts as an effective
hole dopant. The doping dependence of R$_H$ at different
temperatures is shown in the inset of Figure 4. After changing
sign (x$>$0.04), The value of R$_H$(T) initially increases then
decreases at x=0.130. This feature is quite similar to the Hall
effect in Ba$_{0.6}$K$_{0.4}$Fe$_2$As$_2$\cite{ShenBing}, which is
explained as the result of asymmetric scattering between the
electron and hole bands where the mobility is much larger for the
former\cite{FL,ShenBing}.

\begin{figure}
\begin{center}
\includegraphics[width=8cm]{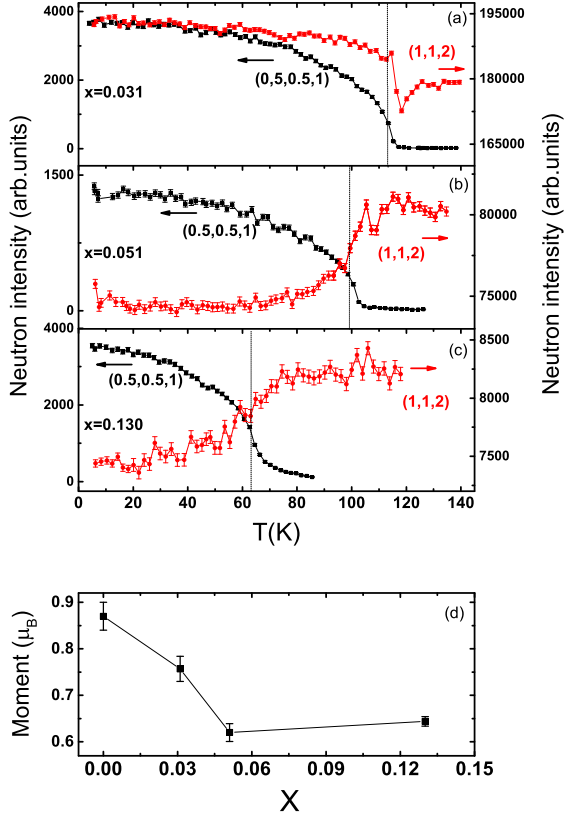}
\caption {Neutron diffraction results for the x=0.031 (a), x=0.051
(b) and x=0.130 (c) single crystals. The temperature dependence of
magnetic Bragg peak (0.5 0.5 1)$_T$ is shown in black squares and
that of nuclear Bragg peak (1 1 2)$_T$ is shown in red circles.
(d) Refined magnetic moments for x = 0, 0.031, 0.051, and 0.130 at
T=5~K. The refinements were carried out using Fullprof software
based on the neutron diffraction data.} \label{Fig.5}
\end{center}
\end{figure}

Neutron diffraction experiments were performed on single crystals
at x= 0.031, 0.051 and 0.130, the results are shown in figure
5(a)-(c). The AFM transition was characterized by measuring the
temperature dependence of (0.5 0.5 1)$_T$ magnetic Bragg peak and
the tetragonal to orthorhombic structural transition is identified
by the intensity change of the (1 1 2)$_T$ nuclear peak (T refers
to tetragonal basis)\cite{WB122}. Combing the neutron scattering
data and resistivity results of d$\rho$/dT, where only one
transition anomaly is observed, both the structural and magnetic
transitions occur at the same temperature for V-doped samples.
This agrees with other hole-doped FeAs-122 cases (Cr, Mn, Mo and
K)\cite{Mn,Cr_Neutron,Mo,WBK}, but different from the
electron-doping cases in which T$_S$ and T$_N$ are well
separated\cite{AQCP}. The doping dependence of the refined
magnetic moment is presented in figure 5(d). For the
electron-doping cases such as Co-doped BaFe$_{2}$As$_{2}$, the
ordered moment is linearly suppressed with doping and finally
becomes zero with 5\% Co-doping\cite{CoMoment}. For the
hole-doping cases like Ba(Fe$_{1-x}$Cr$_{x}$)$_{2}$As$_{2}$, the
magnetic moment is independent of concentration for x$\leq$0.2
despite suppression of the transition temperature from 133 K to 56
K. Then the moment quickly deceases to about 0.3$\mu_B$ at
x=0.335\cite{Cr_Neutron}. While for V-doping case, the moment
evolution seems to be slightly different. The average ordered Fe/V
moment first decreases in the range of 0$\leq$ x $\leq$0.051, then
becomes almost doping independence from x=0.051 (0.62$\mu_B$) to
x=0.13 (0.64$\mu_B$).

\begin{figure}
\begin{center}
\includegraphics[width=8cm]{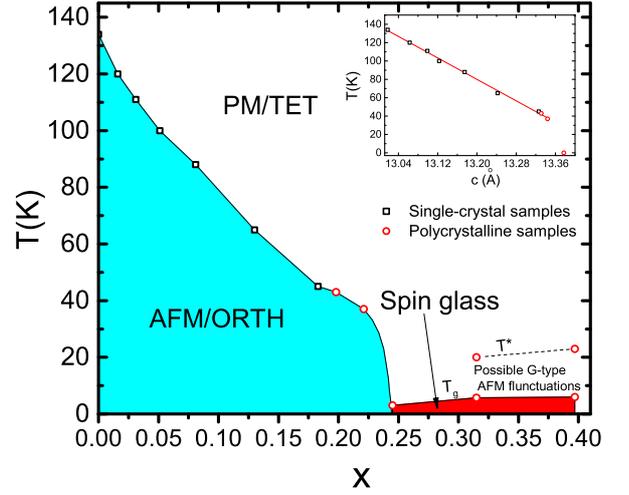}
\caption {Compositional-Temperature phase diagram for
Ba(Fe$_{1-x}$V$_{x}$)$_{2}$As$_{2}$. Spin glass temperature T$_g$
and possible G-type AFM fluctuation temperature T$^*$ are defined
in figure 2. The AFM transition temperature versus $c$ lattice
parameter plot is shown in the inset. The linear fitting result is
shown as the red solid line.} \label{Fig.6}
\end{center}
\end{figure}

Based on the experimental results above, the $T$-$x$ phase diagram
for V-doped BaFe$_2$As$_2$ is presented in figure 6. There is a
gradual suppression of the AFM/structural transition temperature
of BaFe$_2$As$_2$ from 133~K to 37~K with increasing $x$. The
decreasing of AFM transition temperature actually follows a linear
behavior with increasing $c$, see the inset of figure 6. Usually
one would expect an doping-induced magnetic quantum critical
point. However the AFM/structural transition suddenly vanishes at
x=0.245 and $c$=13.54\AA. On the other hand, a spin glass state is
detected on the vanishing point of AFM order. These observations
strongly suggest an avoided quantum critical point (QCP) by the
development of competing states\cite{MnFeSi,AQCP}. Magnetic QCP
should not be a neccessary condition for Fe-based
superconductivity, it is interesting to mention that for Co- or
Ni-doped BaFe$_2$As$_2$, near the optimal doping with maximum
T$_c$, there were also reports about an avoided QCP\cite{AQCP} and
clustered spin glass state\cite{Co_Glass,Ni_Glass}. The spin glass
in BaFe$_{2-x}$(Co/Ni)$_x$As$_2$ was previously interpreted as an
intrinsic response to the competition between the
superconductivity and antiferromagnetism\cite{Co_Glass}, but
similar explanation is not applicable for the SG in
Ba(Fe$_{1-x}$V$_{x}$)$_{2}$As$_{2}$ since no superconductivity was
observed for the V-doped samples. The most plausible explanation
is that the magnetic impurity V in metallic hosts could result in
the Ruderman-Kittel-Kasuya-Yosida (RKKY) effective exchange
interaction between the impurity spins, which leads to a spin
glass behavior\cite{MnRKKY,SG_Review}.

The AFM-like kinks (T*) observed in M-T curves at x=0.315, 0.397
closely resemble the behavior of checkerboard (G-type) AFM order
in Cr- and Mn-doped BaFe$_2$As$_2$\cite{Cr_Neutron,Mn}. According
to calculations in Mn-doped BaFe$_2$As$_2$, the doped magnetic
impurities would exhibit a G-type magnetic structure close to
their cores in heavily doped region when the RKKY interactions
between the conducting electrons and the magnetic impurities are
taken into account\cite{MnRKKY}. Similarly, the T$^*$ in the
V-doped samples most probably represents the formation of
short-range local G-type AFM fluctuations. The Fe-based
superconductivity may be accompanied by the spin fluctuations of
the 3d electrons of Fe which are possibly suppressed by the
competition of G-type spin fluctuations brought by V.

\section{Conclusion}

As a magnetic impurity, the doping of V into BaFe$_2$As$_2$
suppresses the AFM order and generate effective hole-doping
effect, no superconductivity is observed which is in contrast with
the K-doped BaFe$_2$As$_2$. This indicates the extreme sensitivity
of in-plane magnetic impurity for Fe-based superconductivity. On
the other hand, the RKKY spin glass behavior accompanied by
possible checkerboard antiferromagnetic fluctuations emerges at
higher doping levels. The phase diagram of
Ba(Fe$_{1-x}$V$_{x}$)$_{2}$As$_{2}$ exhibits an avoided QCP
similar as that in Ni-doped BaFe$_{2}$As$_{2}$.

\acknowledgments The work at RUC is supported by NSFC (No.11204373
and No.11227906).


\begin{thebibliography}{0}

\bibitem{Cuprate}
  \Name{Lee Patrick A., Nagaosa Naoto, \and Wen Xiao-Gang}
  \REVIEW{Rev. Mod. Phys.}{78}{2006}{17}.

\bibitem{Hosono}
  \Name{Hosono Hideo, Tanabe Keiichi, Takayama-Muromachi Eiji, Kageyama Hiroshi, Yamanaka Shoji, Kumakura Hiroshi, Nahara Minoru, Hiramatsu Hidenori, \and Fujitsu Satoru}
  \REVIEW{Sci. Technol. Adv. Mater.}{16}{2015}{033503}.

\bibitem{FeReview}
  \Name{Bao Wei}
  \REVIEW{Chin. Phys. B}{22}{2013}{087405}.

\bibitem{IC_Ni}
  \Name{Luo Huiqian, Zhang Rui, Laver Mark, Yamani Zahra, Wang Meng, Lu Xingye, Wang Miaoyin, Chen Yanchao, Li Shiliang, Chang Sung, Lynn Jeffrey W., \and Dai Pengcheng}
  \REVIEW{Phys. Rev. Lett.}{108}{2012}{247002}.

\bibitem{Cr}
  \Name{Sefat Athena S., Singh David J., VanBebber Lindsay H., Mozharivskyj Yurij, McGuire Michael A., Jin Rongying, Sales Brian C., Keppens Veerle, \and Mandrus David}
  \REVIEW{Phys. Rev. B}{79}{2009}{224524}.

\bibitem{Cr_Neutron}
  \Name{Marty K., Christianson Andrew D., Wang C. H., Matsuda M., Cao H., VanBebber L. H., Zarestky J. L., Singh David J., Sefat Athena S., \and Lumsden Mark D.}
  \REVIEW{Phys. Rev. B}{83}{2011}{060509(R)}.

\bibitem{Mn}
  \Name{Kim M. G., Kreyssig A., Thaler A., Pratt D. K., Tian W., Zarestky J. L., Green M. A., Bud'ko S. L., Canfield P. C., McQueeney R. J., \and Goldman A. I.}
  \REVIEW{Phys. Rev. B}{82}{2010}{220503(R)}.




\bibitem{Ni_QCP}
  \Name{Zhou R., Li Z., Yang J., Sun D. L., Lin C. T., \and Zheng G. Q.}
  \REVIEW{Nat. Commun.}{4}{2013}{2265}.


\bibitem{C4}
  \Name{Allred J. M., Taddei K. M., Bugaris D. E., Krogstad M. J., Lapidus S. H., Chung D. Y., Claus H., Kanatzidis M. G., Brown D. E., Kang J., Fernandes R. M., Eremin I., Rosenkranz S., Chmaissem O., \and Osborn R.}
  \REVIEW{Nat. Phys.}{12}{2016}{493}.

\bibitem{Co_Glass}
  \Name{Dioguardi A. P., Crocker J., Shockley A. C., Lin C. H., Shirer K. R., Nisson D. M., Lawson M. M., apRoberts-Warren N., Canfield P. C., Bud'ko S. L., Ran S., \and Curro N. J.}
  \REVIEW{Phys. Rev. Lett.}{111}{2013}{207201}.

\bibitem{Ni_Glass}
  \Name{Lu Xingye, Tam David W., Zhang Chenglin, Luo Huiqian, Wang Meng, Zhang Rui, Harriger Leland W., Keller T., Keimer B., Regnault L.-P., Maier Thomas A., \and Dai Pengcheng}
  \REVIEW{Phys. Rev. B}{90}{2014}{024509}.

\bibitem{Co_Sc}
  \Name{Sefat Athena S., Jin Rongying, McGuire Michael A., Sales Brian C., Singh David J., \and Mandrus David}
  \REVIEW{Phys. Rev. Lett.}{101}{2008}{117004}.


\bibitem{NiCu_Sc}
  \Name{Ni N., Thaler A., Yan J. Q., Kracher A., Colombier E., Bud'ko S. L., Canfield P. C., \and Hannahs S. T.}
  \REVIEW{Phys. Rev. B}{82}{2010}{024519}.

\bibitem{RhIrPd_Sc}
  \Name{Han Fei, Zhu Xiyu, Cheng Peng, Mu Gang, Jia Ying, Fang Lei, Wang Yonglei, Luo Huiqian, Zeng Bin, Shen Bing, Shan Lei, Ren Cong, \and Wen Hai-Hu}
  \REVIEW{Phys. Rev. B}{80}{2009}{024506}.


\bibitem{Pt_Sc}
  \Name{Zhu Xiyu, Han Fei, Mu Gang, Tang Jun, Ju Jing, Tanigaki Katsumi, Wen Hai-Hu}
  \REVIEW{Phys. Rev. B}{81}{2010}{104525}.

\bibitem{Ru_Sc}
  \Name{Sharma Shilpam, Bharathi A., Chandra Sharat, Reddy Raghavendra,
Paulraj S., Satya A. T., Sastry V. S., Gupta Ajay, Sundar C. S.}
  \REVIEW{Phys. Rev. B}{81}{2010}{174512}.


\bibitem{P_Sc}
  \Name{Jiang Shuai, Xing Hui, Xuan Guofang, Wang Cao, Ren Zhi, Feng Chunmu, Dai Jianhui, Xu Zhu-an \and Cao Guanghan}
  \REVIEW{J. Phys.: Condens. Matter}{21}{2009}{382203}.

\bibitem{K122}
  \Name{Rotter Marianne, Tegel Marcus, Johrendt Dirk}
  \REVIEW{Phys. Rev. Lett.}{101}{2008}{107006}.

\bibitem{Mo}
  \Name{Sefat Athena S., Marty Karol, Christianson Andrew D., Saparov Bayrammurad, McGuire Michael A., Lumsden Mark D., Tian Wei, \and Sales Brian C.}
  \REVIEW{Phys. Rev. B}{85}{2012}{024503}.



\bibitem{FL}
  \Name{Fang Lei, Luo Huiqian, Cheng Peng, Wang Zhaosheng, Jia Ying, Mu Gang, Shen Bing, Mazin I. I., Shan Lei, Ren Cong, \and Wen Hai-Hu}
  \REVIEW{Phys. Rev. B}{80}{2009}{140508(R)}.


\bibitem{122_Review}
  \Name{Konzen Lance M.N., \and Sefat Athena S.}
  \REVIEW{J. Phys.: Condens. Matter}{29}{2017}{083001}.


\bibitem{Kdoping}
  \Name{Rotter Marianne, Pangerl Michael, Tegel Marcus, \and Johrendt Dirk}
  \REVIEW{Angew. Chem. Int. Ed.}{47}{2008}{7949}.

\bibitem{ShenBing}
  \Name{Shen Bing, Yang Huan, Wang Zhao-Sheng, Han Fei, Zeng Bin, Shan Lei, Ren Cong, \and Wen Hai-Hu}
  \REVIEW{Phys. Rev. B}{84}{2011}{184512}.


\bibitem{WB122}
  \Name{Huang Q., Qiu Y., Bao Wei, Green M. A., Lynn J. W., Gasparovic Y.
C., Wu T., Wu G., \and X. H. Chen}
  \REVIEW{Phys. Rev. Lett.}{101}{2008}{257003}.

\bibitem{WBK}
  \Name{Chen H., Ren Y., Qiu Y., Bao Wei, Liu R. H., Wu G.,
Wu T., Xie Y. L., Wang X. F., Huang Q., \and Chen X. H.}
  \REVIEW{Europhys. Lett.}{85}{2009}{17006}.


\bibitem{AQCP}
  \Name{Lu Xingye, Gretarsson H., Zhang Rui, Liu Xuerong, Luo Huiqian, Tian Wei, Laver Mark, Yamani Z., Kim Young-June, Nevidomskyy A. H., Si Qimiao, \and Dai Pengcheng}
  \REVIEW{Phys. Rev. Lett.}{110}{2013}{257001}.

\bibitem{CoMoment}
  \Name{Lester C., Chu Jiun-Haw, Analytis J. G., Capelli S. C., Erickson A. S., Condron C. L., Toney M. F., Fisher I. R., \and Hayden S. M.}
  \REVIEW{Phys. Rev. B}{79}{2009}{144523}.




\bibitem{MnFeSi}
  \Name{Goko Tatsuo, Arguello Carlos J., Hamann Andreas, Wolf Thomas, Lee Minhyea, Reznik Dmitry, Maisuradze Alexander, Khasanov Rustem, Morenzoni Elvezio, \and Uemura Yasutomo J.}
  \REVIEW{npj Quantum Materials}{2}{2017}{44}.


\bibitem{MnRKKY}
  \Name{Gastiasoro Maria N. \and Andersen Brian M.}
  \REVIEW{Phys. Rev. Lett.}{113}{2014}{067002}.

\bibitem{SG_Review}
  \Name{Binder K. \and Young A. P.}
  \REVIEW{Rev. Mod. Phys.}{58}{1986}{801}.

\end{thebibliography}
\end{document}